\documentclass[useAMS,usenatbib]{mn2e}
\usepackage{graphicx}

\newcommand\aj{{AJ}}%
\newcommand\araa{{ARA\&A}}%
\newcommand\apj{{ApJ}}%
\newcommand\apjl{{ApJ}}%
\newcommand\aap{{A\&A}}%
\newcommand\aaps{\ref@jnl{A\&AS}}%
\newcommand\azh{\ref@jnl{AZh}}%
\newcommand\baas{BAAS}%
\newcommand\caa{\ref@jnl{Chinese Astron. Astrophys.}}%
\newcommand\cjaa{\ref@jnl{Chinese J. Astron. Astrophys.}}%
\newcommand\icarus{Icarus}%
\newcommand\jcap{\ref@jnl{J. Cosmology Astropart. Phys.}}%
\newcommand\jrasc{\ref@jnl{JRASC}}%
\newcommand\memras{MmRAS}%
\newcommand\mnras{{MNRAS}}%
\newcommand\na{\ref@jnl{New A}}%
\newcommand\nar{\ref@jnl{New A Rev.}}%
\newcommand\pra{\ref@jnl{Phys.~Rev.~A}}%
\newcommand\prb{\ref@jnl{Phys.~Rev.~B}}%
\newcommand\prc{\ref@jnl{Phys.~Rev.~C}}%
\newcommand\prd{\ref@jnl{Phys.~Rev.~D}}%
\newcommand\pre{\ref@jnl{Phys.~Rev.~E}}%
\newcommand\prl{\ref@jnl{Phys.~Rev.~Lett.}}%
\newcommand\pasa{\ref@jnl{PASA}}%
\newcommand\pasp{\ref@jnl{PASP}}%
\newcommand\pasj{\ref@jnl{PASJ}}%
\newcommand\qjras{\ref@jnl{QJRAS}}%
\newcommand\skytel{\ref@jnl{S\&T}}%
\newcommand\solphys{\ref@jnl{Sol.~Phys.}}%
\newcommand\sovast{\ref@jnl{Soviet~Ast.}}%
\newcommand\ssr{Space~Sci.~Rev.}%
\newcommand\zap{\ref@jnl{ZAp}}%
\newcommand\nat{Nature}%
\newcommand\iaucirc{\ref@jnl{IAU~Circ.}}%
\newcommand\aplett{\ref@jnl{Astrophys.~Lett.}}%
\newcommand\apspr{\ref@jnl{Astrophys.~Space~Phys.~Res.}}%
\newcommand\bain{\ref@jnl{Bull.~Astron.~Inst.~Netherlands}}%
\newcommand\fcp{\ref@jnl{Fund.~Cosmic~Phys.}}%
\newcommand\gca{\ref@jnl{Geochim.~Cosmochim.~Acta}}%
\newcommand\grl{\ref@jnl{Geophys.~Res.~Lett.}}%
\newcommand\jcp{\ref@jnl{J.~Chem.~Phys.}}%
\newcommand\jgr{\ref@jnl{J.~Geophys.~Res.}}%
\newcommand\jqsrt{\ref@jnl{J.~Quant.~Spec.~Radiat.~Transf.}}%
\newcommand\memsai{\ref@jnl{Mem.~Soc.~Astron.~Italiana}}%
\newcommand\nphysa{\ref@jnl{Nucl.~Phys.~A}}%
\newcommand\physrep{\ref@jnl{Phys.~Rep.}}%
\newcommand\physscr{\ref@jnl{Phys.~Scr}}%
\newcommand\planss{\ref@jnl{Planet.~Space~Sci.}}%
\newcommand\procspie{\ref@jnl{Proc.~SPIE}}%
          
\title[Chaotic Dynamics of 1P/Halley]{Chaotic Dynamics of Comet 1P/Halley; \\
    Lyapunov Exponent and Survival Time Expectancy}
\author[M. A. Mu\~noz-Guti\'errez et. al.]{M. A. Mu\~noz-Guti\'errez$^{1}$\thanks{E-mail:
mmunoz@astro.unam.mx}, M. Reyes-Ruiz$^{2}$ and B. Pichardo$^{1}$\\
$^{1}$Instituto de Astronom\'ia, Universidad Nacional Aut\'onoma de M\'exico, Apdo. postal 70-264 Ciudad Universitaria, M\'exico\\
$^{2}$Instituto de Astronom\'ia, Universidad Nacional Aut\'onoma de M\'exico, Apdo. postal 877, 22800 Ensenada, M\'exico}
\begin{document}

\date{Accepted 1988 December 15. Received 1988 December 14; in original form 1988 October 11}

\pagerange{\pageref{firstpage}--\pageref{lastpage}} \pubyear{2002}

\maketitle

\label{firstpage}

\begin{abstract}

   The orbital elements of comet Halley are known to a very high precision, 
    suggesting that the calculation of its future dynamical evolution is straightforward.
In this paper we seek to characterize the chaotic nature of the present day orbit of comet Halley and to quantify the timescale over which its motion can be predicted 
     confidently. In addition, we attempt to determine the timescale over which its present day 
     orbit will remain stable. Numerical simulations of the dynamics of test particles in orbits similar to that of comet
    Halley are carried out with the Mercury 6.2 code. On the basis of these we construct 
    survival time maps to assess the absolute stability of Halley's orbit, frequency analysis maps,
    to study the variability of the orbit and we calculate the Lyapunov exponent for the 
    orbit for variations in initial conditions at the level of the present day uncertainties in our 
    knowledge of its orbital parameters. On the basis of our calculations of the Lyapunov exponent for comet Halley, the
    chaotic nature of its motion is demonstrated. The {\it e}-folding timescale for the divergence of initially
    very similar orbits is approximately 70 years. The sensitivity of the dynamics on initial conditions is
    also evident in the self-similarity character of the survival time and frequency analysis maps
    in the vicinity of Halley's orbit, which indicates that, on average, it is unstable on a timescale of 
    hundreds of thousands of years. The chaotic nature of Halley's present day orbit implies that a precise determination of 
    its motion, at the level of the present day observational uncertainty, is difficult to predict on a 
    timescale of approximately 100 years. Furthermore, we also find that the ejection of Halley
    from the solar system or its collision with another body could occur on a timescale as short as 10,000 years.
    
\end{abstract}


\section{Introduction}

Halley's comet is probably one of the most studied and therefore best
known minor bodies in the Solar System to date. Historical records of comet Halley
start in the year 240 B.C. \citep{Kiang72}, but it is until its last
perihelion passage, in February 1986, when it became visible to modern
telescopes and even physically accesible to spacecrafts, that the
amount of available data has hugely grown. In particular, the
parameters of its retrograde orbit, semimajor axis, $a$, and
eccentricity, $e$, are since then determined with a precision of the
order of $10^{-6}$ \citep[$\sigma_q=771\times10^{-9}$ and $\sigma_e=91\times10^{-8}$ respectively, where $\sigma$ is standard deviation and $q$ the perihelion distance,
according to][]{Landgraf86}.

The origin of Halley's comet has been a matter of discussion for
decades. One of the likely sources of Halley-type comets (i.e.
short-period comets with Tisserand parameters $T < 2$ with respect to
Jupiter, periods $20 < P < 200$ years and semimajor axes less than 40 AU),
seems to be the Oort cloud \citep{Fernandez80}. Indeed, since giant
planet perturbations of trans-Neptunian objects in the vicinity of the Kuiper Belt, will not generally
drive comets to the observed inclinations on Halley-type comets, so the
origin of their orbits must be different. In their
computations, \citet{Fernandez80}, and also \citet{Duncan88}
and \citet{Quinn90}, show that the dynamical evolution of comets from the Oort cloud
toward the inner solar system, is a probable origin of the random
inclination of Halley-type comets, since they tend to preserve their
random orbital inclinations. On the other hand, \citet{Levison01},
noticed that the inclination distribution of Halley-type comets is not
isotropic, meaning that they could not be readily explained as
originating from a rather isotropic source as the Oort cloud, but from a flattened component or inner disc-like portion of the Oort cloud. 
Earlier, \citet{Duncan97} had
investigated a viable origin of Jupiter family comets coming from the already known
flattened scattered disc \citep{Luu97}. They found that
approximately 1\% of their scattered disc objects survived the full 4 Gyr simulation, where some of these reach semimajor axis of thousands of AU.
In a more
recent paper, \citet{Levison06} show that these objects, once they reach a semimajor axis of the order of $10^4$ AU are rapidly reduced in their perihelia due to galactic tides. If just 0.01\% of these comets then evolve, due to giant planet interactions, onto Halley-type orbits, the resultant statistical orbital distribution is consistent with observations.

The physical properties of comet Halley are also known as never before
\citep[for a review see][]{Mumma11}. \citet{Ahearn95} have shown that
Halley-type comets differ from Jupiter Family comets on their average
coma carbon abundances, suggesting a different origin for both
families. Comet Halley seems to lose mass at an approximate rate of
0.5\% every perihelion passage \citep{Whipple51,Kresak87}. At this
rate, the comet might be severely diminished or even vanished in about
15,000 years. Halley's comet has been spreading particles that settle
down on the known Orionid stream for thousands of years
\citep{Sekhar13}.

From the dynamical point of view the evolution of this comet has also
been profusely studied through numerical integrations
\citep{Yeomans81,Dvorak90,Levison94,Bailey96,VdHelm12,Sekhar13}. Detailed
calculations show for example that Halley's comet has been trapped in
the past by resonances with Jupiter and secular perturbations, such as
the Kozai resonance and other secular resonances \citep{Quinn90,Bailey96,Thomas96}
affecting considerably its long term dynamical
evolution \citep{Sekhar13}. From numerical simulations under various approximations, Halley's comet and other Halley-type comets seems to be intrinsically chaotic \citep{Petrosky86,Froeschle88,Chirikov89}.

In the pioneering work of \citet{Chirikov89} they used a simple
analytical model based on the results of integrations by
\citet{Yeomans81} which resembles the perihelion passage over the past
$\sim3000$ years, in order to obtain a measure of the chaotic behavior
of the Halley's comet. They found that Jupiter plays the major role in
driving the local instabilities of the motion while Saturn contributes
an order of magnitude less in the random changes suffered by the
comet. Apart from these early work, not many attempts have been made to obtain a reliable quantification of the chaotic
behavior of Halley's comet with modern numerical integration
tools. This means that a standard Lyapunov exponent or Lyapunov time
has not been established with confidence for Halley by means of
direct numerical integrations.

In this work we explore numerically the evolution of test particles
in the surrounding phase space of Halley's comet in order to determine
the chaoticity of this region quantified through frequency analysis
maps and a simple auxiliary visual tool that we have called ``survival
time maps''. Additionally we compute, for the first time directly from numerical
integration, the Lyapunov exponent for Halley's comet in its
current orbit considering its observational uncertainty. By finding a
positive exponent we have determined that current known orbit is
actually chaotic. We also estimated the {\it e}-folding time scale for the separation of neighbouring orbits. Finally we provide an estimation of sojourn
time in the solar system for the comet in a qualitative similar manner to
that used by \citet{Chirikov89} finding a median value almost an order of magnitude smaller than theirs.

The outline of this paper is as follows: in Section 2 we describe the
three different analysis techniques to determine the chaotic nature of
the dynamics of Halley's comet and the corresponding numerical simulations
done for this purpose. In Section 3 we show results of the numerical
simulations while discussing these results in Section 4. Finally we
give our conclusions in Section 5.

\begin{figure*}
\includegraphics[width=18cm]{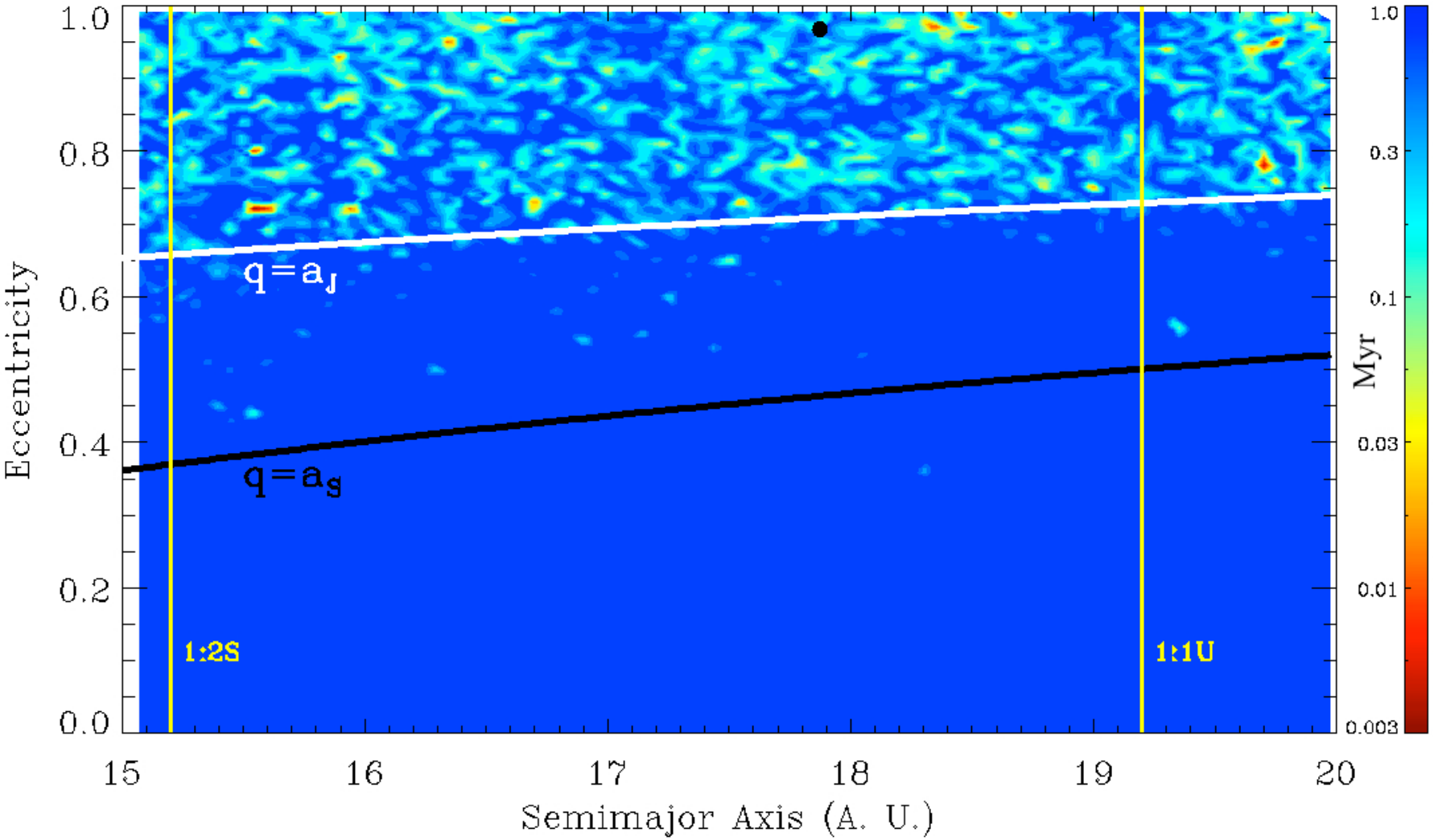}
\caption{Survival Time Map for orbits with semimajor axis up to a few AU 
         from comet Halley's orbit, all with the same inclination. 
         The color scale indicates the survival time as a function of initial 
         semimajor axis and eccentricity of the orbit. Halley's present 
         day orbit is indicated by the black circle. The thick  
         black and white lines correspond to particles crossing the orbit of 
         Saturn and Jupiter, respectively. Also shown with vertical yellow
         lines are the strongest mean motion resonances in the region, 1:2
         with Saturn and 1:1 with Uranus \citep{Gallardo06}.\label{STM_zoom1}}
\end{figure*}


\section{Methods and Simulations}

We have used 3 different analysis methods to determine the chaotic nature of Halley's comet dynamics. We describe these in the next subsections.

\subsection{Survival time maps}

Survival time maps (STM hereafter) are an auxiliary tool to
visualize the absolute stability, against ejection from the system or collision with other bodies in orbits of a given phase space region, corresponding in our case to the plane of semimajor axis, $a$, {\it vs} eccentricity, $e$, surrounding Halley's current position in this plane. In the STM a color code indicates the total survival time in the simulation according to the particle's initial condition given by its position in the phase space plane. In order to explore this we have used the MERCURY integrator package developed by \citet{Chambers99} to integrate the orbits of clones of comet Halley as test particles in a Solar System N-body simulation. The clones were generated in such a way that their initial conditions cover a region on the $a-e$ plane with different zoom levels. 

We performed 3 different simulations to construct STMs corresponding to different ranges of $a$ and $e$. In the first simulation a) semimajor axis, $a$, ranges from 15 to 20 AU and $e$ ranges from 0 to 1 with both intervals divided uniformly in 100 values each, giving a total of $10^4$ particles. In the second simulation b) $a$ ranges from 17.385 to 18.385 AU divided uniformly in 100 values and $e$ ranges from 0 to 1 for a total of 3300 particles (33 values in $e$). Although the small $e$ space comprises a very different dynamical region than that for comet Halley, we explore it to put into perspective the importance of giant planets influence zones and to get a glimpse into the possible behavior of other Halley-type comets that may be explored in a future work. In the third simulation c) the domain of $a-e$ space covers the observational error according to \citet{Landgraf86} both in $a$ ($\sigma_a\sim10^{-6}$ AU) and $e$ ($\sigma_e\sim10^{-6}$) sampled with $10^4$ total particles. In all cases the angular elements, argument of pericentre, $\omega$, longitude of the ascending node, $\Omega$, mean anomaly, $M$ and inclination, $i$, were set the same as that of Halley's comet as given by the Horizons web site\footnote{http://ssd.jpl.nasa.gov/horizons.cgi} at the beginning date of the simulations, Oct. 1, 2012 in heliocentric coordinates. A final simulation d) was performed varying semimajor axis and inclination and letting the 4 remaining orbital parameters, including eccentricity, be the same as for comet Halley on the same date, and covering the observational error in $a-i$ phase space 
\citep[with $\sigma_i\sim10^{-5}$ deg, according to][]{Landgraf86}. We divide both ranges in 100 uniformly spaced values for a total of $10^4$ particles.

For a) and b) simulations the RADAU integrator from \citet{Everhart85} as implemented in the MERCURY package was selected for the integrations with a tolerated accuracy parameter of $10^{-10}$ and an initial time step of 20 days. Simulations c) and d) were carried on using the optimized Bulirsch-St\"oer integrator implemented in the MERCURY package (BS2) with a tolerated accuracy parameter of $10^{-12}$ and an initial time step of 1 day, in order to be confident in the integration of small perihelion, high eccentricity orbits. All simulations spanned over $10^6$ years and include as N-bodies the planets except Mercury in order to avoid incorporating relativistic effects. We also include 5 Dwarf Planets (all but Sedna which orbits far beyond the 100 AU sojourn limit imposed in the simulation) and 5 of the greatest minor bodies (Orcus, Quaoar, Varuna, Ixion and 2002 AW$_{197}$) not yet classified as dwarf planets but likely to be in the near future.

According to the sojourn criterion for STM, the particle is still part of the simulation at the end of the $1\times10^6$ years of it, regardless the changes in orbital parameters of each particle. So a particle is lost when its $a$ becomes greater than 100 AU, collides with the Sun or with another planet.

\subsection{Frequency analysis map}

We have used the frequency analysis introduced by \citet{Laskar93,Laskar90} to quantify the chaotic behavior of particles around current position of Halley's comet. According to \citet{Correia05,Morbidelli02,Laskar93}, the difference $(D)$ in the value of the fundamental frequency of the motion $(n)$ of a particle under consideration, obtained over two consecutive and equal time intervals $(T)$, is a measure of the secular stability of its trajectory and a reliable indicator of chaos. In this context we have explored the $D$ parameter, or diffusion parameter, for clone particles covering a grid of $a$ vs $e$ phase space. For each particle in the map we calculate using a lomb-scargle analysis \citep{Scargle80} the mean frequencies of the mean longitude of the particle, $\lambda(t)=M(t)+\omega(t)+\Omega(t)$, in the adjacent time intervals from 0 to $2.5\times10^5$ yr and from here to $5\times10^5$ yr. The mean motion $n$ in each interval is defined as the amplitude of the mean frequency of the series over $2\pi$ in that interval and then, according to the definition the $D$ parameter is calculated as:
\begin{equation}\label{Dparam}
D=\frac{|n-n'|}{T}
\end{equation}
If $D$ is close to zero, this is if $n\approx n'$, then it means that the orbit is stable. Otherwise high values of $D$ reflect important changes in the motion of the particle related to unstable or even chaotic dynamics.   

We then performed one $5\times10^5$ yr simulation using the RADAU integrator from  the MERCURY package with a toleration accuracy parameter of $10^{-10}$ and an initial time step of 20 days. We cover a semimajor axis span of just 1 AU from 17.385 to 18.385 AU, this is $\pm0.5$ AU from the current semimajor axis of comet Halley, and from 0 to 1 in $e$. We have used 100 uniformly spaced particles in $a$ and 33 particles to uniformly cover $e$, in order to save computational time as we are bounded to record an extensive output cadence of parameters evolution.

\subsection{Lyapunov Exponent}

We have calculated the Lyapunov exponent for 6 neighbouring orbits around the current orbit for comet 
Halley separated by the present day observational uncertainty. The 6 orbits are the result of varying initial
position by $\pm 10^{-6}$ AU in each cartesian axis {X,Y,Z} in a heliocentric reference frame, while 
maintaining the velocity at the same value of the fiducial orbit. This means that each one of the 6 orbits 
differs initially from the fiducial one by a distance of $10^{-6}$ AU distance, $\delta(0)$.

According to \citet{Morbidelli02} in a numerical experiment one can define a small initial separation distance 
between an arbitrary orbit and the fiducial one; the modified orbit is characterized by the vectors 
$\delta\textbf{q}(0)$ and $\delta\textbf{p}(0)$, where $\textbf{q}$ and $\textbf{p}$ are generalized 
coordinates in the problem. One can compute their evolution, $\delta\textbf{q}(t)$, $\delta\textbf{p}(t)$ 
over some constant time interval $T$ after which one can measure the separation distance of the 
modified orbit from the fiducial one and define the quantity:

\begin{equation}
s_j=\Vert\delta\textbf{q}(T),\delta\textbf{p}(T)\Vert/\Vert\delta\textbf{q}(0),\delta\textbf{p}(0)\Vert
\end{equation}

\noindent which we use to determine a new initial separation for the modified orbit in the next step of 
the integration as $\delta_1\textbf{q}(0)=\delta\textbf{q}(T)/s_j$ and $\delta_1\textbf{p}(0)=\delta\textbf{p}(T)/s_j$. 
In this manner, as was demonstrated by \citet{Benettin80}, the Lyapunov exponent, $\mathcal{L}$, can 
be calculated in an iterative process according to:

\begin{equation}
\mathcal{L}=\lim_{l \to +\infty}\frac{\sum_{j=1}^{l}\ln s_j}{lT}
\end{equation}

\noindent where $\mathcal{L}$ is independent of the choice of $T$.  In our case, as we are using only an initial 
separation in distance, i.e. in $\textbf{q}$, it follows that $\delta_i\textbf{q}(0)=\delta(0)=10^{-6}$ AU for all our $i$ steps. 

In this context we have used the Bulirsch-St\"oer integrator from the MERCURY package in order to perform 
a set of 30 simulations, extending over 100 years each in succession, of Halley's fiducial orbit plus 6 orbits
around it. The total simulated time is 3000 years, starting from Oct. 1st, 2012, in order to avoid the first 
close encounter of the fiducial Halley with Jupiter expected to occur at about $\sim$ 3400 years into the future. These close encounters produce random variations in the orbit that 
are capable of modifying the separation of the orbits by several AU on a very short timescale (see figure \ref{separation} 
for details). The initial time step was chosen to be 3 days with an accuracy parameter of $10^{-12}$ in order to 
obtain high enough precision to calculate a real separation of the orbits, resulting from dynamical effects, and 
not a computational artifact. In this manner we have calculated the Lyapunov exponent for each one of the 6 orbits 
previously defined using the iterative process just as described, and a maximum Lyapunov exponent for the full set of orbits.

\begin{figure}
\includegraphics[width=\hsize]{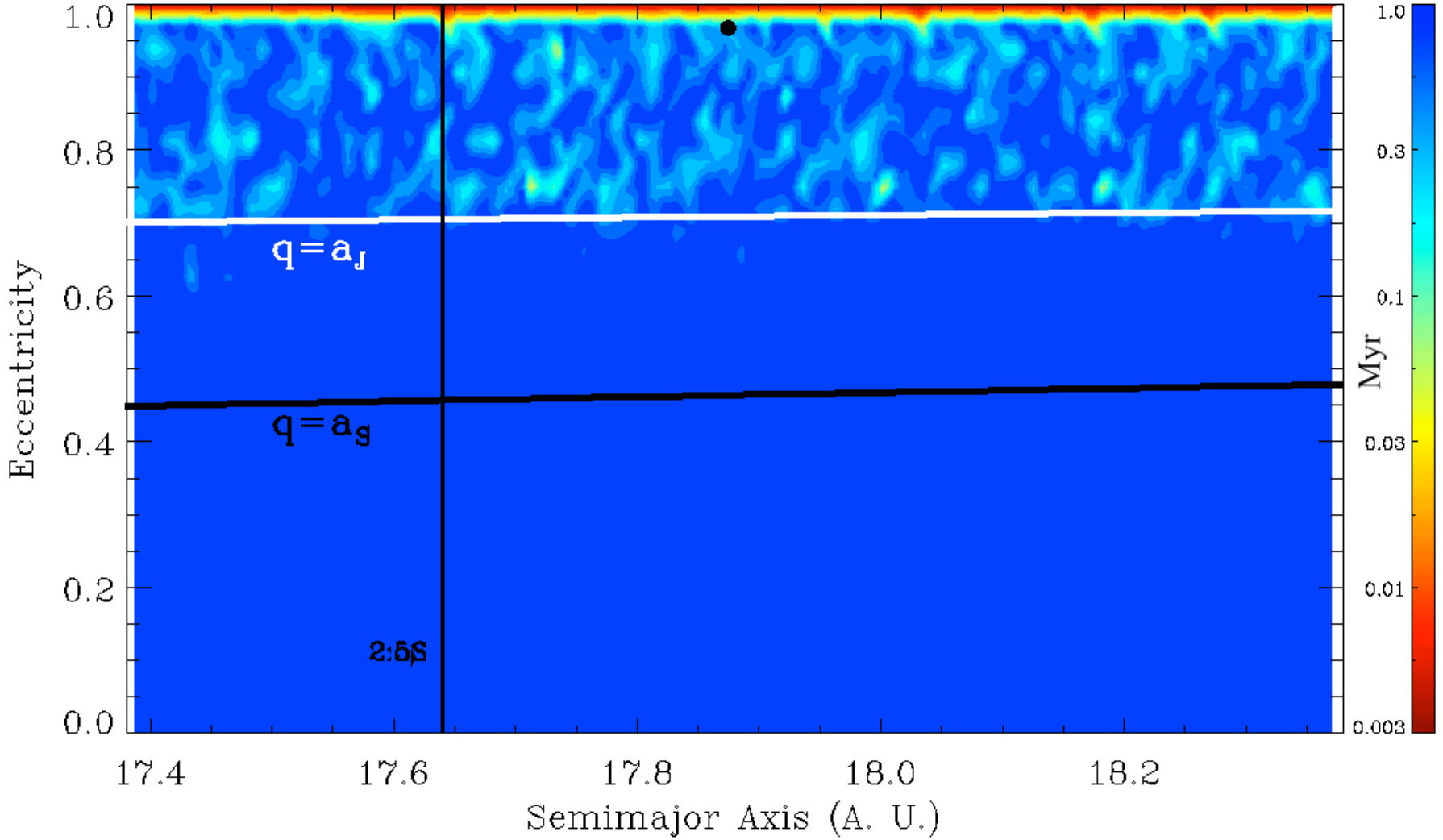}
\caption{Same as Figure \ref{STM_zoom1} but zooming-in on the orbits with 
         semimajor axis within 0.5 AU of Halley's comet orbit. The stronger
         MMR in this region, the 2:5 resonance with Saturn, is indicated by the vertical black line. 
         \label{STM_zoom2}}
\end{figure}

\section{Results}

In order to assess the stability of the orbit of comet Halley (and possibly other
Halley-type comets), and to characterize its dynamics, we have carried out 
numerical simulations as previously described. The results of these studies are
presented in this section.  

\begin{figure*}
\includegraphics[width=18cm]{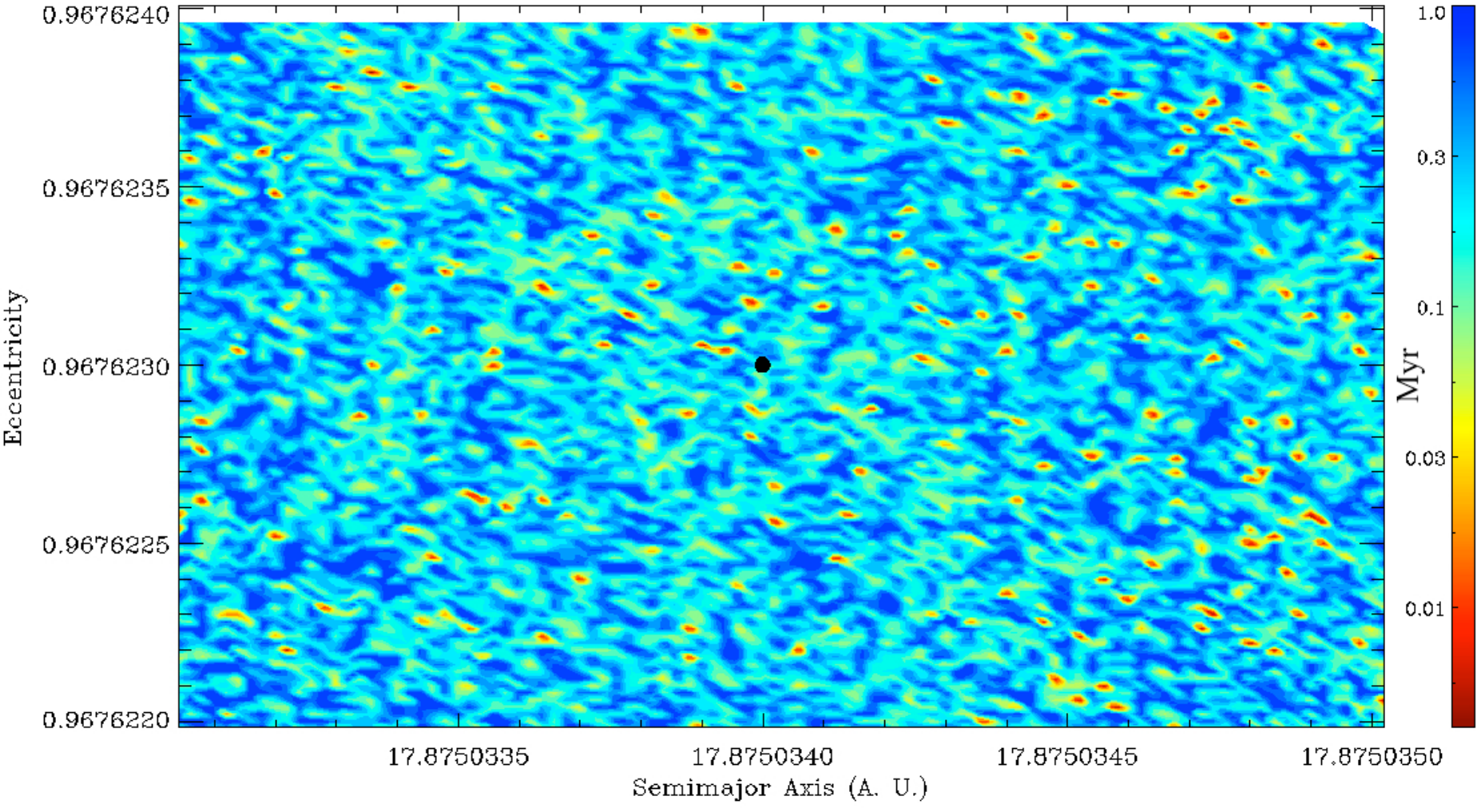}
\caption{Same as Figure \ref{STM_zoom1} but zooming-in on the range of orbital
         parameters ($a-e$) defined by the present day observational uncertainty 
         in Halley's comet orbit according to \citet{Landgraf86}.
         \label{STM_zoom3_ae}}
\end{figure*}

\subsection{Survival Time Maps} \label{RSTM}

Figure \ref{STM_zoom1} shows the survival time map (STM) for orbits with initial semimajor
axis extending from 15 to 20 AU and eccentricity less than 1. The 
test particles sampling these orbits all have the same inclination, corresponding to
the present day inclination of comet Halley, 162.18042 $\deg$ according to the 
Horizons website for the beginning date of our simulations. 
For comparison, we also show in the figure the nominal orbit of comet Halley (black dot). 

Evident in Figure \ref{STM_zoom1} is the fact that a wide range of orbits crossing that of 
Jupiter are strongly unstable, with a median lifetime, for such orbits, of $5.6\times10^5$ years. 
Note that this lifetime is a lower limit for the expected lifetime of small bodies in this region 
of $a-e$ phase space, as the survival time of stable orbits may be much greater than the 1 Myr timescale 
considered in our simulations. This result is also seen in Figure \ref{STM_zoom2}
which shows the STM zooming-in on the semimajor axis to a range extending within 
0.5 AU of Halley's orbit. In this case the median survival time is $5.1\times10^5$ years which 
is consistent with the result from the previous map, considering the use of less particles on a broader range in eccentricity.

An important feature steaming from the comparison of the sequence of increasing zoom Figures
\ref{STM_zoom1}, \ref{STM_zoom2} and \ref{STM_zoom3_ae}, the last one corresponding to 
a semimajor axis and eccentricity range covered by the observational uncertainty for comet 
Halley \citep{Landgraf86} is the fact that the ``structure'' of the stable/unstable zones is 
apparently self-similar likely fractal in nature. The determination of the fractal dimension for these 
structures is beyond the scope of the present study. 
This result is already suggestive of strongly chaotic dynamics for small bodies in the region, 
as orbits within a very small neighborhood in $a-e$ space can have extremely different behaviour 
regarding their stability properties.   

\begin{figure}
\includegraphics[width=\hsize]{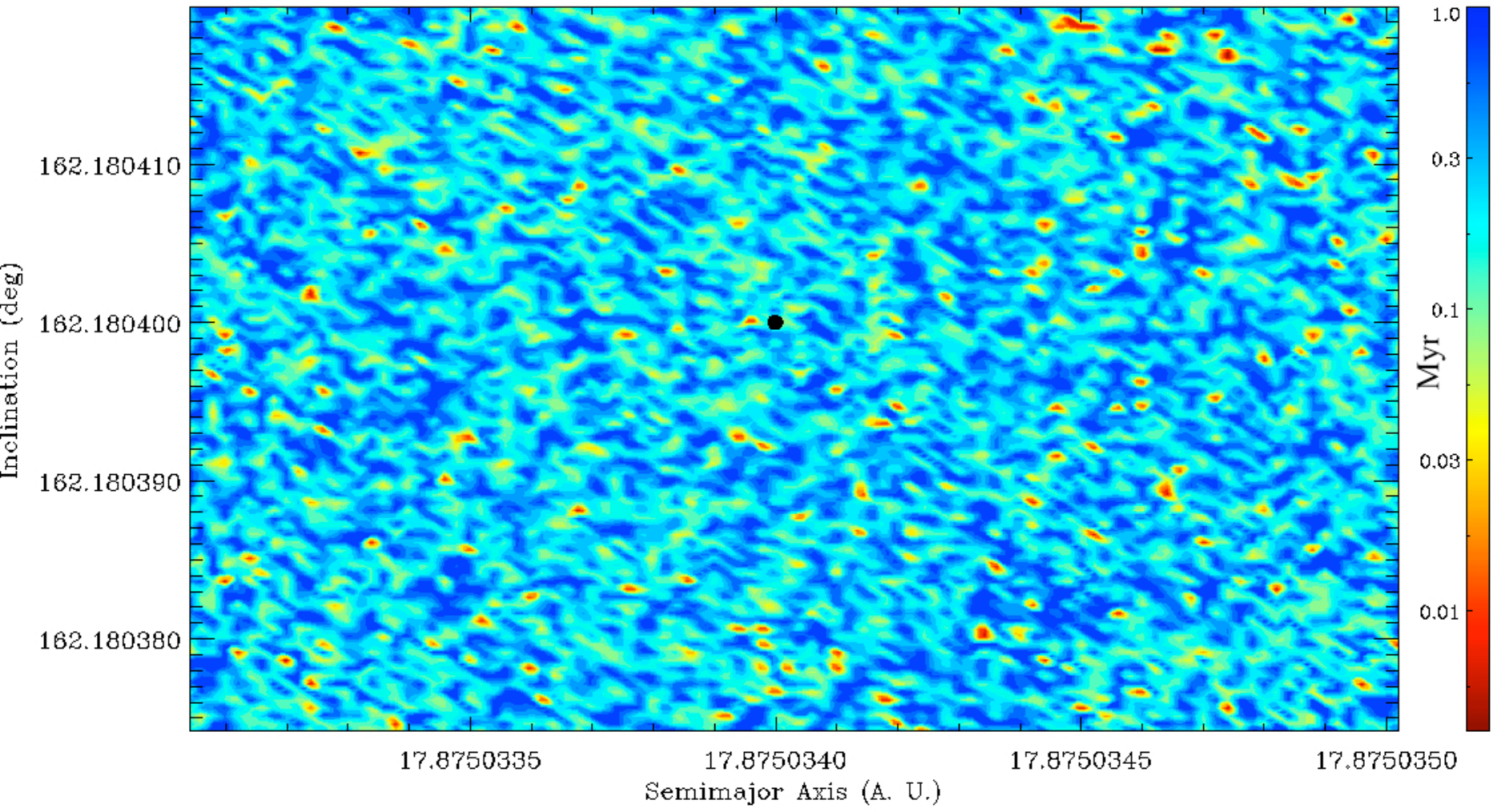}
\caption{Survival Time Map for orbits with semimajor axis and inclination 
         within the observational uncertainty of comet Halley \citep{Landgraf86}. The 
         eccentricity in all cases is taken at the nominal value.
         \label{STM_zoom3_ai}}
\end{figure}

A similar result is found in the structure of
the stable/unstable zones, where we plot them as a function of orbit inclination of the test particles. Figure \ref{STM_zoom3_ai}
shows the STM for orbits with the present day eccentricity of comet Halley, 0.967623, 
again according to the Horizons website, but different inclinations spanning over the $10^{-5}\deg$ 
observational uncertainty in this parameter \citep{Landgraf86}.

In both $a-e$ and $a-i$ phase space stability maps extending over the observational uncertainty 
(figures \ref{STM_zoom3_ae} and \ref{STM_zoom3_ai}) the median survival time for particles in 
the region is found to be $3.2\times10^5$ yr. This value is then a reliable estimate of 
the stability time for the current known orbit of comet Halley in the Solar System.

\subsection{Frequency analysis maps} \label{RFAM}

\begin{figure*}
\includegraphics[width=18cm]{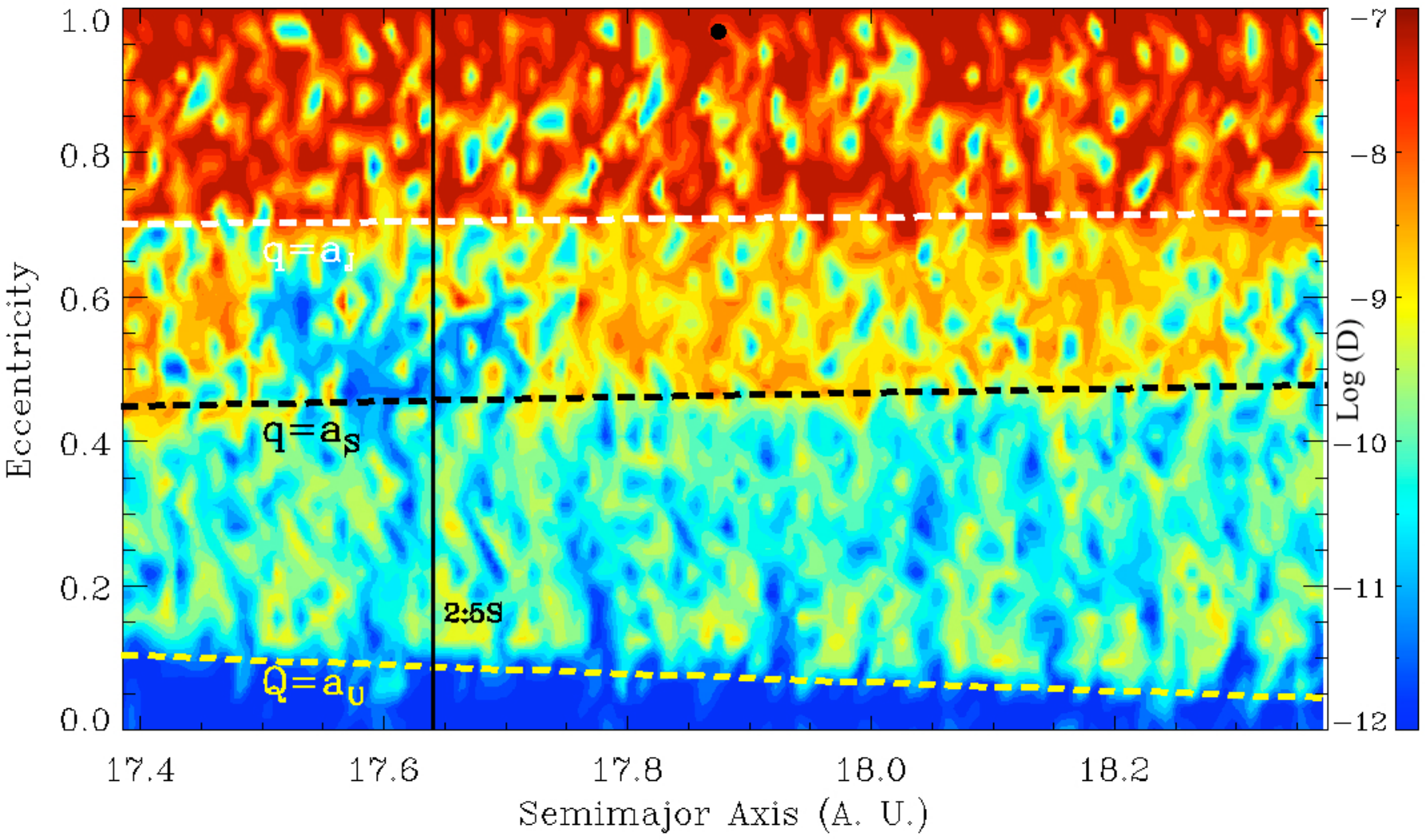}
\caption{Frequency analysis map for orbits with comet Halley's inclination but differing
         in semimajor axis (by up to 0.5 AU) and eccentricity. Red colored regions correspond 
         to the most unstable orbits. Horizontal lines indicate 
         orbits with perihelia crossing Jupiter's orbit (white dashed), Saturn's orbit 
         (black dashed) and with aphelia crossing the orbit of Uranus (yellow dashed). 
         The black vertical line indicates the position of the 2:5 MMR with Saturn, the stronger 
         one in the region \citep{Gallardo06}.
         \label{LASKAR}}
\end{figure*}

A more detailed analysis of the stability of orbits within 0.5 AU of Halley's orbit (with its 
nominal inclination) is shown in Figure \ref{LASKAR}, where contours of the rate of change 
in the mean motion of particles initially on a given orbit, over the extent of a $5\times10^5$ year 
simulation, are plotted. Red and orange coloured areas in this figure denote regions of phase 
space where orbital parameters, particularly the mean motion, change by more than 5\% 
over this timescale. While these orbits are not unstable in the sense of the results presented in 
the STMs of section \ref{RSTM}, they do change significantly and may lead to a different 
dynamical evolution of particles on these zones. 

As in the STM corresponding to the same zoom level (Figure \ref{STM_zoom2}), a region of 
strong orbital variation is shown in Figure \ref{LASKAR} for orbits with perihelia that cross the
orbit of Jupiter, indicated by the (nearly) horizontal, dashed white line. A large proportion of 
orbits above this line exhibit significant variation over the length of the simulation. In addition, 
a pair of families of unstable orbits are also found, corresponding to those particles that cross 
the orbit of Saturn and do not cross that of Jupiter, and to those  particles with aphelia large 
enough to cross the orbit of Uranus but not to cross Saturn's orbit. These features are 
not found on the basis of the STMs previously presented, as they do not correspond to strictly
unstable orbits. A mean value for the $D$ parameter in each region is found to be as follows: 
$D\sim10^{-7.8}$ for particles crossing Jupiter's orbit, $D\sim10^{-9.2}$ for particles crossing 
Saturn's orbit but not Jupiter's, and $D\sim10^{-10.1}$ for particles that reach the orbit of 
Uranus but not Saturn's.

A simple numerical estimation of the time scale for orbits to evolve significantly according to the 
$D$ parameter, $\tau_D$, can be found from the definition of diffusion parameter given in 
equation \ref{Dparam}, from which it is clear that $\tau_D \sim n/D\sim 1/PD$ years, where $P$ is 
the orbital period. As $P\sim75$ years across this whole region of phase space, the corresponding 
time scale in the three regions mentioned above is $\sim 8 \times 10^5$, $\sim 2 \times 10^7$ and 
$\sim 2 \times 10^8$ years, respectively. The timescale $\tau_D$ can be understood as the time 
required for the particle to suffer major changes in its orbit. The instability time 
obtained from STM and Laskar frequency analysis is roughly consistent for particles crossing 
Jupiter's orbit. In addition, these estimations are also consistent with the fact that particles not 
crossing Jupiter's orbit can survive the whole 1 Myr simulation, as found from the STMs.

Islands of stability/instability (blue/red zones) exist throughout the region mapped in Figure \ref{LASKAR}, 
where no relation with major resonances or 
other bodies in the solar system are easily identified.
Although we have not carried out simulations 
to accomplish the Laskar frequency analysis at different zoom levels, the rapidly changing structure 
of the stable/unstable regions of phase space suggests a strong dependence on the initial conditions 
for the dynamical evolution of small bodies in the region. 

\subsection{Lyapunov Exponent} \label{Rlyexp}

The strong sensitivity on initial conditions depicted by our results in connection to the STMs and the 
Laskar stability analysis, is further illustrated in Figure \ref{separation} which shows the   
separation distance between 2 orbits and the nominal solution for comet Halley. One of the orbits 
considered (red line in Figure \ref{separation}) is for a particle starting from initial coordinates 
with the $X$ component of its position greater than the nominal solution by an amount $\delta_0 = 10^{-6}$ AU, 
the reported observational error for the ephemerides of comet Halley. The other orbit differs initially
from Halley's nominal solution in the same amount but in the $Y$ coordinate (blue line in Figure 
\ref{separation}). Both orbits rapidly diverge from the nominal solution. 

As illustrated in Figure \ref{separation}, there are 2 processes leading to an increase in the separation 
of neighbouring orbits. A gradual increase, as that occurring up to 3,400 yr during which the 
separation increases by approximately 5 orders of magnitude due to the effect of distant encounters 
with other bodies in the system. And a phase of abrupt change in orbital parameters due to a 
close encounter of the particle with Jupiter. During this phase, the dynamical evolution is particularly sensitive
to the initial conditions, as can be seen from the widely different separation for the orbits considered
following the close encounters. These events, during which the particle approaches the planet 
to within 3 Hill radii ($R_{\rm H} = (M_p / 3 M_{\odot})^{1/3}$), are marked by black vertical 
lines in Figure \ref{separation}, a bit after 3,400 yr and at slightly less than 3,800 yr.      

The formal definition of the term chaos is matter of debate. In this paper we follow 
\citet{Strogatz94} who proposes a working definition of chaotic behaviour as ``aperiodic long-term 
behaviour in a deterministic system that exhibits sensitive dependence on initial conditions.'' Our 
system is clearly deterministic and the sensitive dependence on initial conditions is usually 
measured in terms of the Lyapunov exponents which quantify the exponential rates at which 
neighbouring orbits diverge (or converge) as the system evolves in time. If such system exhibits 
at least one positive Lyapunov exponent, then it is said that the dynamics of the system is chaotic.

\begin{figure}
\includegraphics[width=\hsize]{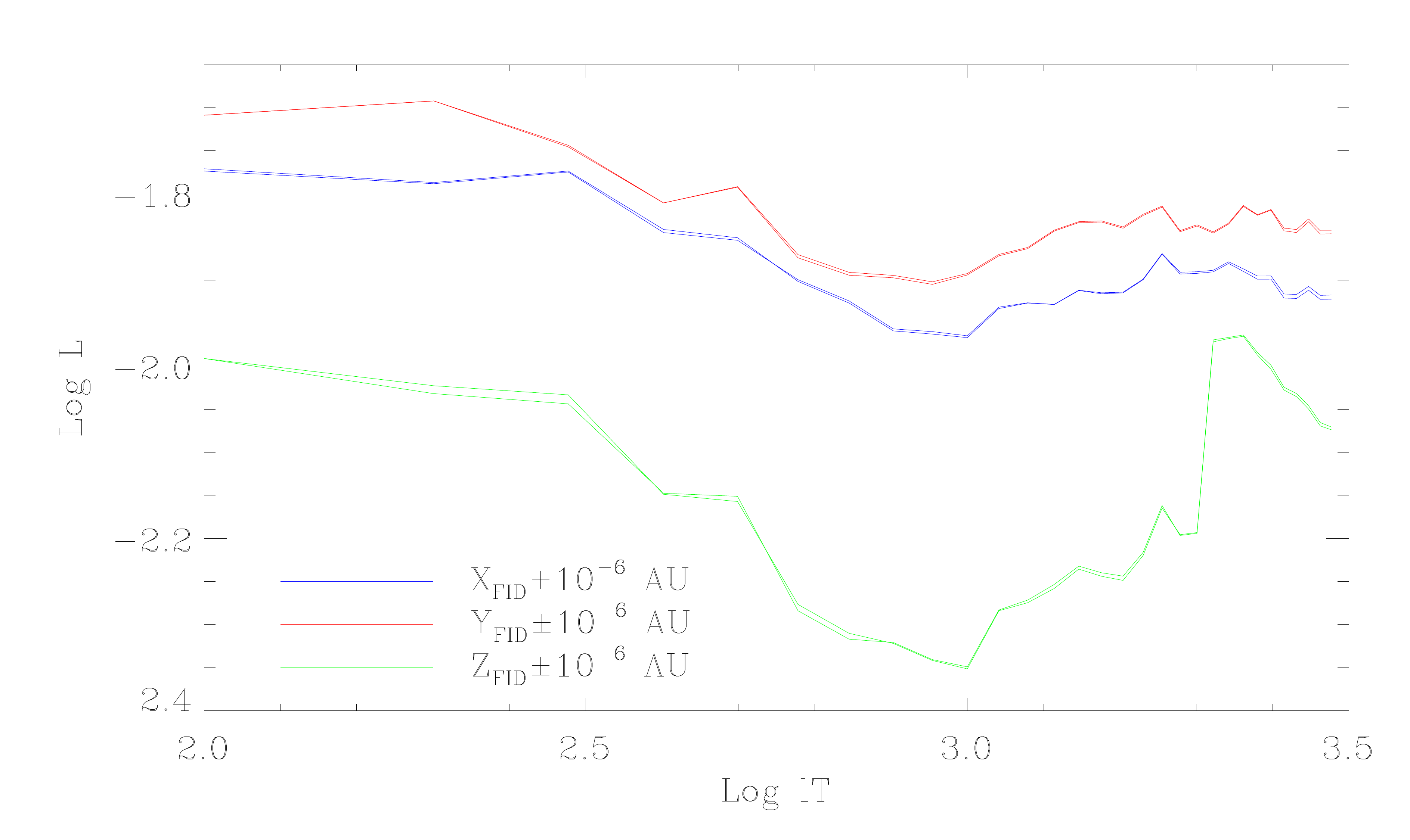}
\caption{Lyapunov exponents for the 6 orbits after 3000 years integrations. The maximum Lyapunov exponent which results from variations in the Y cartesian axis (red curves) implies a Lyapunov time of $\sim70$ years.\label{lyapunov}}
\end{figure}

To formally assess the chaotic behaviour of comet Halley we have computed the maximum Lyapunov 
exponent for particles on Halley's nominal orbit, with respect to variations in the orbital parameters at the 
level of their current observational uncertainty. The resulting Lyapunov exponent is shown in 
Figure \ref{lyapunov}. The corresponding Lyapunov timescale is approximately 70 yr. It is important 
to point out that this timescale corresponds to the time scale for orbits to diverge by 1 part in
$10^6$, which is approximately the observational uncertainty. 
 
\section{Discussion}

In this section we discuss in more detail some of the implications of our results and analyse the effect of 
some of the assumptions and methodology used in our study.  

\subsection{Predictability of Halley's orbit}

One of the important consequences of the chaotic nature of Halley's orbit is  
the possibility of making long term predictions of its motion. 
On one hand, we can interpret our results of section \ref{Rlyexp} in terms of the 
so-called Lyapunov timescale, $\tau_{\mathcal{L}}$, defined as the inverse 
of $\mathcal{L}$, corresponding to the e-folding timescale of the separation of initially 
neighbouring orbits. The fact that  $\tau_{\mathcal{L}}\approx 70$ yr, is an indication that over such timescale two orbits initially separated in semimajor axis, eccentricity 
or inclination by 1 part in 10$^6$, diverge from one another 
by a factor of $e$ in separation. This precludes the possibility of making high precision 
predictions, at a level similar to that of the present-day observational uncertainty in its orbital 
parameters, the ``short'' term, $\sim 70$ yr, evolution of the orbit of comet Halley, i.e. from one
perihelion passage to the next.

\begin{figure}
\includegraphics[width=\hsize]{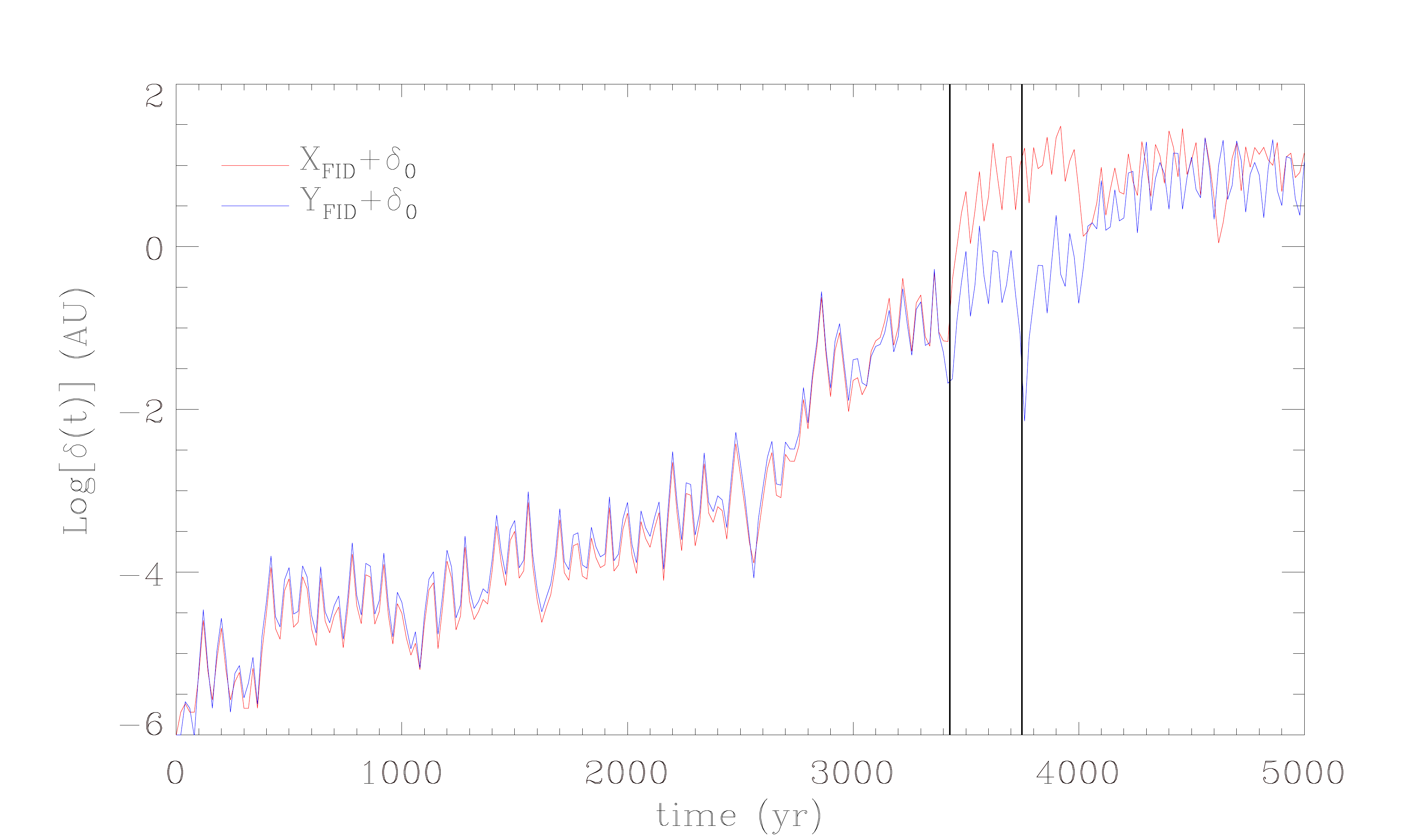}
\caption{Evolution of the separation distance between modified orbits and 
         Halley's fiducial orbit. Curves correspond to initial conditions in which the 
         $X$ coordinate (red line) and the $Y$ coordinate (blue line) are varied by $\delta_0$ 
         from the fiducial coordinate for Halley's comet. Vertical black lines indicate
         times of close encounters between the comet on the fiducial orbit and Jupiter.
         \label{separation}}
\end{figure}

On the other hand, the fact that the contours in the survival time maps, which reflect the 
structure of regions of dynamical stability in $a-e$ phase space, have an 
apparently self-similar or fractal nature (commonly found in chaotic dynamical systems) 
as can be seen from comparing Figures \ref{STM_zoom1}, \ref{STM_zoom2} and \ref{STM_zoom3_ae} corresponding to STMs at
different ``zoom'' levels around Halley's orbit, also suggests that the survival time of the comet 
can not be accurately determined. As evident in Figure \ref{STM_zoom3_ae}, closely neighbouring orbits 
(within the observational uncertainty) can have widely different stability properties. 
The time over which Halley remains stable can range from 10$^4$ to 10$^6$ yr (or more), 
with the median value in a domain ranging over the observational uncertainty of $a$ and $e$
being approximately 3.2 $\times$ 10$^5$ yr, at the zoom level shown in Figure \ref{STM_zoom3_ae}. 
Furthermore, this result implies that an orbit, such as Halley's nominal solution, apparently located
in a stability island on a given ``zoom'' level, may turn out to be much more unstable as we 
zoom-in around it. A similar conclusion is reached from the distribution of values 
for the Laskar $D$ index which measures the change in the orbit and has also an
apparent ``fractal'' structure as shown in Figure \ref{LASKAR}. 

\begin{figure}
\includegraphics[width=\hsize]{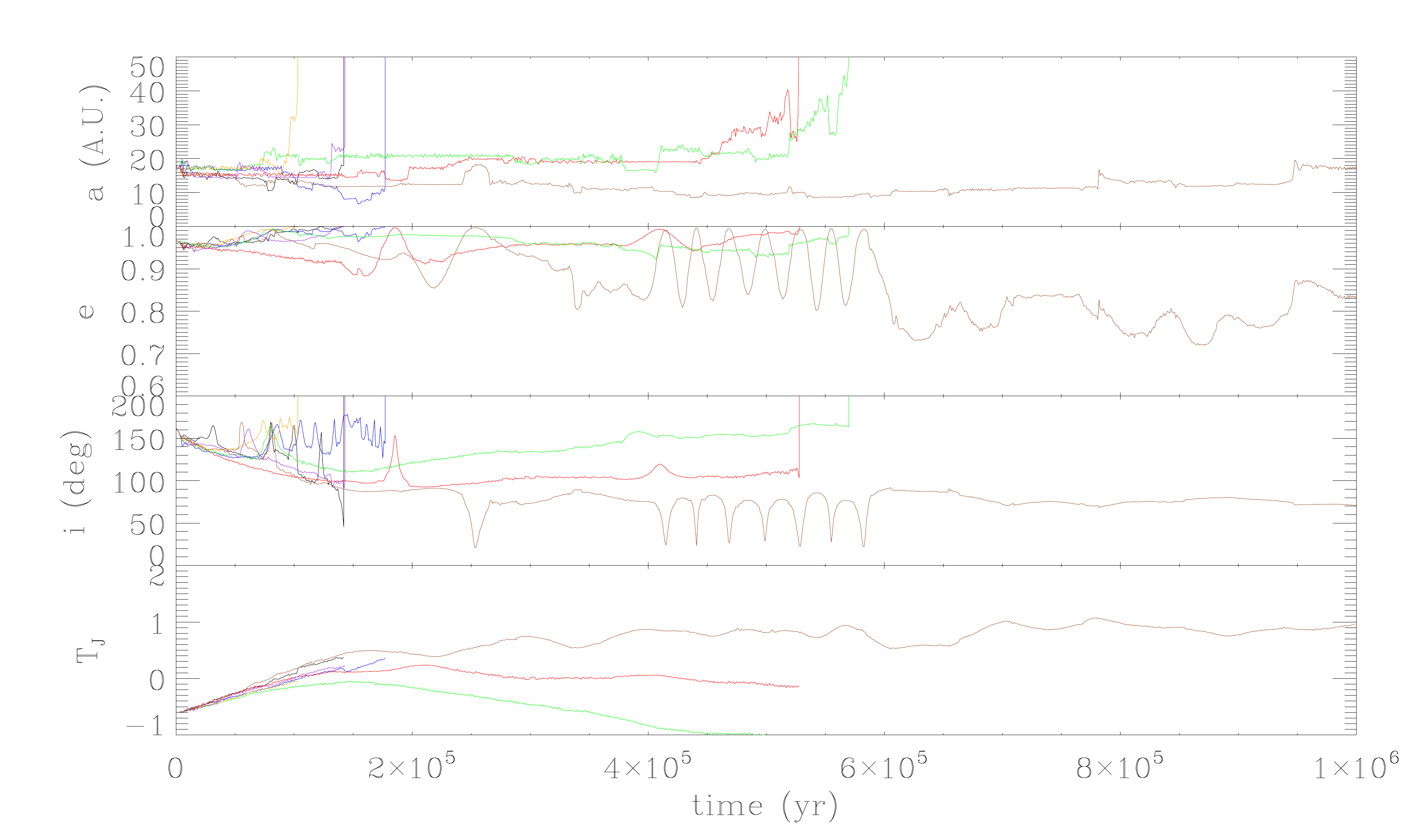}
\caption{Temporal evolution of the main orbital parameters of 7 particles with initial 
  conditions within the present day observational uncertainties in comet Halley's orbit. Particles 1 and 2 
  correspond to initial conditions deviating in $\pm 10^{-6}$ AU in the $X$ coordinate from the 
  center of the $a-e$ box defined in Figure \ref{STM_zoom3_ae}, similarly, particles 3 and 4 and 5 and 6 
  correspond to initial conditions within $\pm 10^{-6}$ AU in the $Y$ and $Z$ coordinate, respectively.
  Particle 7 starts out from initial conditions corresponding to the center of the box. 
  \label{aeiTvst_7}}
\end{figure}

Another illustration of the strong dependence of Halley's motion on its initial position and 
velocity, and of the difficulty in predicting its future motion, is given in Figure \ref{aeiTvst_7},
which shows the evolution of the orbital parameters $a$, $e$, $i$ and the Tisserand parameter 
with respect to Jupiter, $T_J$, for 7 particles located at extreme values of these parameters, 
within the box defined by the observational uncertainty for Halley's orbit. Over a timescale of 
$10^5$ years or more, even orbits starting out with a difference in orbital parameters of less 
than the observational uncertainty, can have widely different outcomes.

\subsection{On the fate of Halley's orbit}

Halley-type comets as a class (HTCs hereafter), are characterized by orbital periods less than 200 yr 
and Tisserand parameter with respect to Jupiter smaller than 2. 
In the present study we 
have assumed that comet Halley is already on its present day orbit and followed its subsequent 
evolution, not much can be said about the origin of its orbit. 

\begin{figure}
\includegraphics[width=\hsize]{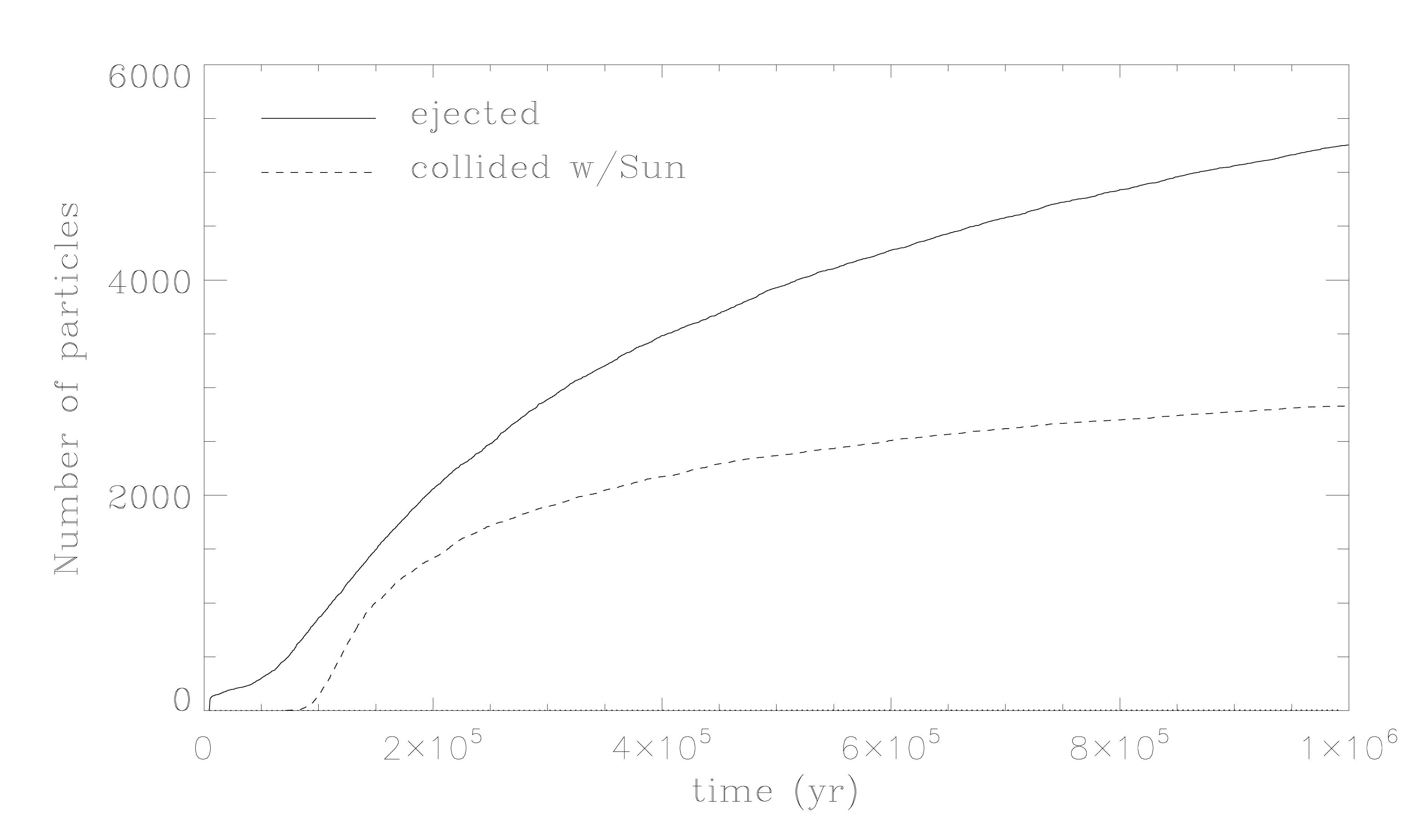}
\caption{Temporal evolution of the number of particles ejected from the solar system and
that collide with the Sun out of the $10^4$ particles originally present in the simulation. All particles
are started from a set of orbits consistent with the present day orbit of comet Halley, i.e with
$a$ and $e$ values within the observational uncertainties. Additionally, from the whole set of particles, 6 collided with one of the planets.
  \label{PartEnd_ObsErr}}
\end{figure}

Nevertheless, the results presented in Section \ref{RSTM} indicate that the survival time for 
objects in Halley-type orbits ranges from 10$^4$ to 10$^6$ yr, implying that a body in a Halley-like
orbit can remain on it, or on one very similar to it, for no more than 
approximately 1 Myr. This, together with the fact that Halley is still an active comet, which
suggests that it has probably been in its present, short period and small perihelion orbit, 
less than 10$^4$ yr, implies that Halley will probably remain on it for at least a similar 
timescale. 

In order to estimate the probability that comet Halley remains in a stable orbit similar to the 
one it has in the present day, we have computed the fraction of particles remaining in the simulations 
as a function of time. Figure \ref{PartEnd_ObsErr} shows the number of particles colliding 
with the Sun or ejected from the solar system, for a collection of 10,000 particles started from
initial conditions within a domain in $a-e$ phase space defined by the observational 
uncertainties in these parameters, as we did in the construction of the STM of 
Figure \ref{STM_zoom3_ae}. The initial sharp increase in the number of ejected particles is related with the first close encounter of all particles with Jupiter after $\sim$ 3400 yr of the beginning of the simulation. It's worth to mention that although all particles suffer this close encounter, as a result of it just a few of them are ejected from the system. This due likely to the chaotic nature of its motion. We find that $10^5$ yr after the start of the simulation, 2\% of
all particles have collided with the Sun and approximately 10\% have been ejected from the system. 
Hence, we could estimate that comet Halley has approximately a 90\% chance of surviving 
at least $10^5$ yr in the solar system. However, in a similar manner we can estimate that 
by $10^6$ yr, there is a 28\%  chance that it will collide with the Sun and a 50\% chance 
that it will be ejected from the system. 

\begin{figure*}
\includegraphics[width=\hsize]{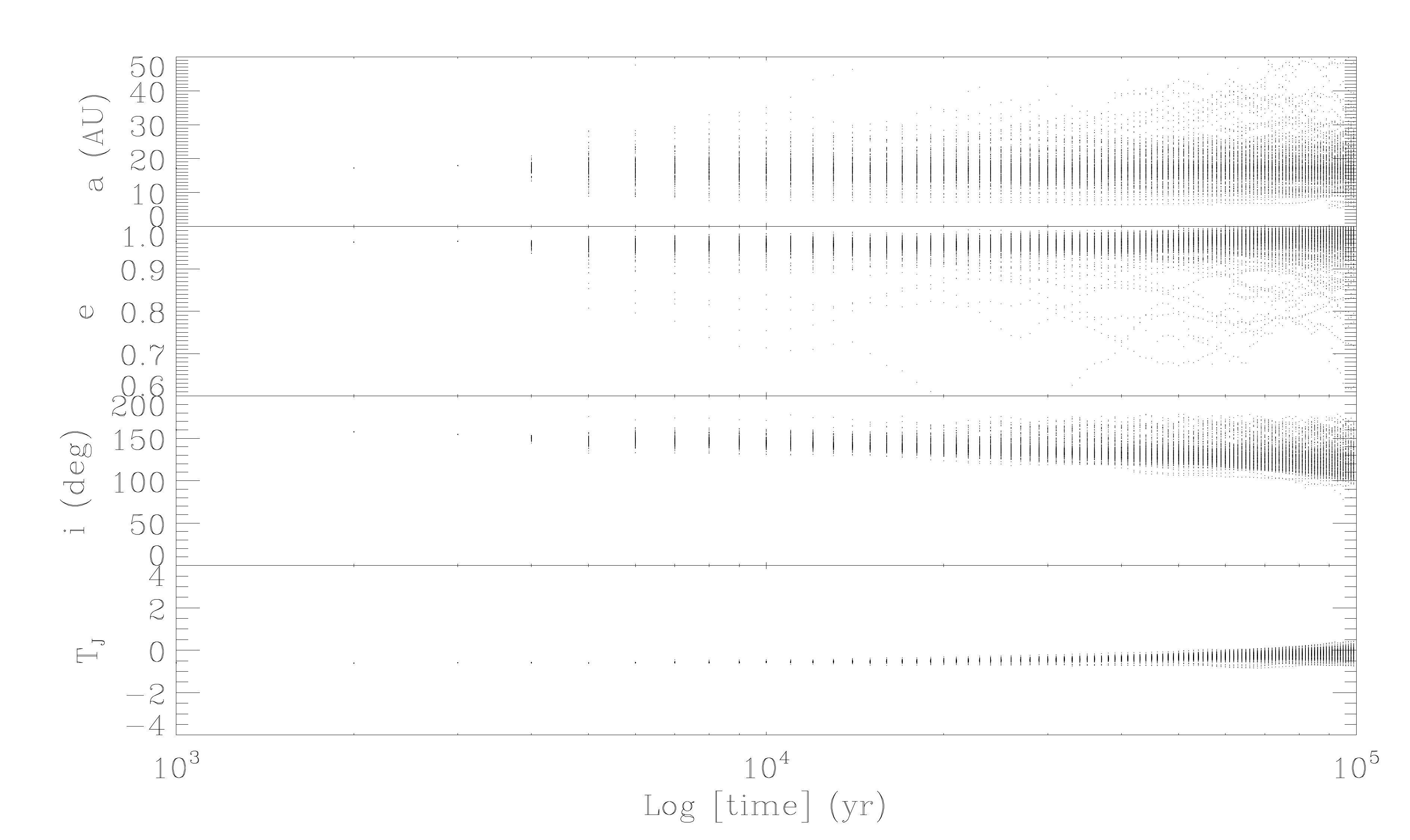}
\caption{Temporal evolution of the main orbital parameters of 200 randomly chosen particles with initial 
  conditions within the present day observational uncertainties in comet Halley's orbit. All particles
  have initial conditions corresponding to orbits consistent with the present day orbit of comet 
  Halley. Each point represent the instantaneous value of the parameters for a particular solution at a given time. 
  \label{aeiTvst_200}}
\end{figure*}

One final piece of interesting information to be gathered from our results is that the orbits 
consistent with comet Halley (within the observational uncertainty) that are not lost from 
the system due to ejection or collision with the Sun, have a tendency to evolve conserving a
Tisserand parameter with respect to Jupiter, as seen in Figure \ref{aeiTvst_200}, which 
shows the evolution of $a$, $e$, $i$ and $T_J$ for 200 randomly chosen particles in our 
simulations. Again, the initial orbit of all these particles is consistent with Halley's present 
day orbit, and most of the particles in the ensemble are seen to evolve as follows:

\begin{itemize}
\item A rapid increase, on a timescale of a few thousand years, in the spread of $a$, $e$ and 
$i$ representing the diversity of possible orbits to be followed by Halley if it survives.
\item A large fraction of orbits evolve into periods greater than 200 years, i.e. they are no
longer short period comets. 
\item The average eccentricity of the ensemble increases slightly on a timescale of $10^5$ yr.
\item The inclination of most possible orbits consistent with Halley's current orbit tends 
to evolve into lower inclination, almost polar orbits.
\end{itemize} 

Although there is a wide spread in possible outcomes for comet Halley as we have discussed 
in previous sections, it seems likely that, if it survives for more than $10^5$ yr, its orbit will 
evolve into a more eccentric and less inclined orbit.  

\subsection{Origin of chaos}

It is generally believed that chaotic dynamics of small bodies in the solar system, 
such as asteroids in the Main Belt or comets in the Kuiper Belt, results from the 
overlapping of mean-motion and secular resonances with the major bodies in the 
solar system \citep[for a review of the topic see][an references therein]{Malhotra98}. 

Comet Halley however, is not trapped in any of the known 
strong resonances with the outer planets, and this is not expected due to the comet's 
high eccentricity and inclination. So the origin of the chaotic character of its orbit 
is not straightforward to identify. Furthermore, it appears that the chaotic motion we
have identified is not directly related to the close encounters of the comet with the 
giant planets, particularly with Jupiter. This is a well known source of strong chaos 
in the system which however, develops on a greater timescale of many orbital periods.
This is illustrated in Figure \ref{separation}, where the first close encounters of the 
comets with Jupiter are shown to occur after approximately 50 orbital periods. 

The weak chaotic behaviour we have characterized by the Lyapunov exponent 
analysis of section \ref{Rlyexp} is not directly related to these close encounters, as 
it develops long before the first close encounter occurs. One possible explanation for the origin of the chaotic motion of Halley's comet is the overlap of $p:1$ mean motion resonances, where $p$ is an integer, with the binary conformed by the Sun and Jupiter \citep{Shevchenko14}. In a previous work \citet{Shevchenko07} performs an analytical estimation of the Lyapunov time for the Halley's comet obtaining a lower limit for this of $\sim$ 34 yr, which is consistent with our numerical estimation. We intend to analyse this issues in
more detail in a future work.

\subsection{Neglected dynamical effects}

In the present study we have neglected the effect of non-gravitational forces known
to affect the dynamics of active comets near perihelion \citep{Marsden68}. The effect 
on comet Halley is particularly well known, but along with this is the fact that the estimated 
time in which the gaseous jet forces are actively present in comets, is just a few hundred 
perihelion passages, which corresponds to a few thousand years of the dynamical 
life-time of Halley-type comets. As we are interested in the long-term dynamical evolution 
of Halley's comet, we decided to neglect this effect in the simulations 
presented in this work.
 
To test the previous assumption, we carried out a simulation of the long-term 
evolution of Halley's comet including non-gravitational forces as prescribed in the Horizons website\footnote{http://ssd.jpl.nasa.gov/horizons.cgi}, to construct a STM for the conditions studied in Figure \ref{STM_zoom1}. 
In doing that we are highly overestimating the importance of  this effect because we implicitly 
assume that the comet is active for the whole 1 Myr simulation. Nevertheless, even in this 
extreme scenario, the results as far as survival expectancies are statistically equivalent 
to the one without non-gravitational forces presented in the previous sections. 

Finally, it is worth mentioning that the results we have obtained are strictly related with the gravitational 
interaction of test particles with the major solar system bodies, except for Mercury. The mass of 
Mercury could be added to that of the sun, but this adds just $\sim$ 1.6 $\times\ 10^{-7}$ M$_\odot$ 
to the total mass of the Sun, therefore, in the long-term, there is not a measurable difference for Halley's 
comet fate. To correctly account for the influence of Mercury in a simulation of the solar system, it would 
require a relativistic treatment of the equations of motion, but given the chaotic nature of the problem 
and the short time-scale of the simulations, compared to the solar system life-time, we considered 
this a second order effect in our analysis to be considered in future studies. 
        
\section{Conclusions.}

We have carried out a series of numerical simulations aimed to assessing the 
dynamical stability of Halley's comet. Three types of analysis are carried out 
to demonstrate and characterize the chaotic behaviour of the Halley's orbit on the 
basis of Survival Time Maps, Frequency Analysis Maps and a direct calculation
of the Lyapunov exponent. 
 
From our analysis of Survival Time Maps we conclude that it is common 
for Halley-like orbits to be unstable up to the point of being ejected from the system, 
or colliding with another solar system body. We also find that the long term evolution 
of comet Halley, even with the high precision of its observed orbital parameters,  
is difficult to predict on account of the strong dependence on the initial conditions. 
The timescale for particles in orbits differing in less than today's observational  
uncertainties in $a$ or $e$, can range by at least 2 orders of magnitude, from 
10$^4$ to $10^6$ yr approximately. The median timescale for the survival of 
particles in a range of $a$ and $e$ within the observational constraints is 
$3.2 \times 10^5$ yr. 

Based on the Laskar stability analysis in a neighborhood of Halley's comet, we conclude 
that even stable orbits, in which the particles are not ejected or collide with a
solar system body, can change significantly on a timescale of millions of years. 
Again, the precise determination of what will be the fate of comet Halley is hindered
by the strong dependence on the initial conditions, even within today's observational 
uncertainties at the level of 1 part in 10$^6$. On timescales of more than $10^5$ years,
it seems likely that the Halley, if it does not collide with the Sun or is ejected from the 
solar system, will evolve into a higher eccentricity, lower inclination orbit.  

Finally, we have computed the Lyapunov exponent for the present day Halley's
orbit and a series of other orbits differing in $a$ and $e$ by an amount equivalent 
to the observational uncertainty. We have found that ${\mathcal{L}}$ is greater than 
zero with a value of approximately $10^{-2}$, indicating that the orbit is indeed 
chaotic. The corresponding timescale for the prediction of Halley's orbit to
within present day observational constraints is less than 100 years, suggesting 
that the orbit of Halley's comet can not be accurately predicted for timescales
much greater than this. An important finding in our work is that the chaotic behaviour 
is not related to close encounters of Halley with any of the planets in the solar system,
nor to the overlap of any known system of resonances. The origin of the chaos in such 
eccentric orbits is a subject to be explored in more detail in future studies.   

 \section*{Acknowledgments}

     The authors acknowledge Carlos Ch\'avez for his help in the 
      implementation of the routines for the Frequency Analysis Maps. We are also very grateful to the anonymous referee for comments that improved the quality of this paper. We
      also acknowledge support from DGAPA-UNAM PAPIIT grants
      No. IN115109 and No. IN114114. MAM is thankful to CONACYT-Mexico
      for a scholarship to conduct graduate studies.

\bsp

\label{lastpage}

\end{document}